\documentclass[12pt,tightenlines,eqsecnum,floats,aps,amsmath,amssymb,nofootinbib,prd]{revtex4}

\usepackage{setspace}
\usepackage{graphicx}
\usepackage{appendix}
\usepackage{amsmath,amssymb,amsfonts,amsthm,mathrsfs}
\usepackage{enumerate} 
\usepackage[usenames,dvipsnames]{xcolor}
\usepackage{hyperref} 
\hypersetup{colorlinks=true, citecolor=MidnightBlue, linkcolor=MidnightBlue, urlcolor=MidnightBlue}

\begin{document}

\title{Analysis of improved dynamics of non-rotating charged black holes}

\author{Florencia Benítez$^1$, Esteban Mato$^1$ and  Javier Olmedo$^2$}
\affiliation{
1. Instituto de F\'{\i}sica, Facultad de Ingenier\'{\i}a (Universidad de la República), Julio Herrera y Reissig 565,
11300 Montevideo, Uruguay.\\
2. Departamento de Física Teórica y del Cosmos, Universidad de Granada, Granada-18071, Spain.}

\begin{abstract}
We provide the quantization of a charged black hole. We consider a redefinition of the scalar constraint in order to render the algebra of constraints as a Lie algebra. We apply loop quantum gravity techniques adhered to a novel improved dynamics scheme. We show that the model is solvable in closed form.  We compute effective geo\-me\-tries, and show that the resulting effective space-times replace the inner horizon with a transition surface that connects trapped and antitrapped regions within the charged black hole interior. Quantum effects therefore stabilize the classical inner Cauchy horizons, as long as the charge is small  compared to the mass. We further discuss the properties of these effective geometries by defining an effective stress-energy tensor out of the Einstein tensor, concluding that the null energy condition is violated.  
\end{abstract}
\maketitle

\section{Introduction}

Black holes are ultra compact objects that have become a  trending topic of research nowadays \cite{gw16,m87}, and the natural focal point where we expect to find deviations from classical general relativity. They are traditionally thought to be a generic consequence of the gravitational collapse of classical matter \cite{penrs}. It involves the formation of a trapped region and, ultimately, a singularity, where the classical theory breaks down. The first solution to the Einstein equations, describing the formation of a black hole was obtained by Oppenheimer and Snyder \cite{opp-sny}. It entails the collapse of a perfectly spherical distribution of a homogeneous and pressureless perfect fluid (or dust cloud). The eventual outcome of this collapse is a Schwarzschild black hole, characterized by a spacelike curvature singularity enclosed by a trapped region.

Initially, the status of the black hole singularity has remained ambiguous, raising questions about whether it was an inherent outcome or a byproduct of spherical symmetry. But it was soon established as a generic result through the celebrated singularity theorems (see for instance \cite{haw-ell}). These theorems dictate that, once a trapped region forms and specific conditions regarding energy positivity and causality are met, the collapse process inevitably goes on until a singularity forms (in the sense of the inextendability of incomplete geodesics). Notably, these theorems do not provide insights into the nature of the singularity or the causal structure enveloping it.

Actually, the internal structure of realistic black holes seems to be notably more intricate than that portrayed by the Oppenheimer-Snyder solution. For instance, perturbations away from spherical symmetry result in spacelike singularities with distinctive characteristics compared to their symmetric counterparts. This is the case if one adds a small charge or rotation to the Schwarzschild solution, resulting in a geometry with a singularity structure that significantly departs from unperturbed one. For example, the Kerr-Newman solution \cite{kerr-new} features not only an outer apparent horizon but also an inner one. Given that the trapped region does not extend all the way to the singularity, the nature of this singularity adopts a timelike character. 
Moreover, they show intriguing causal features \cite{kerr-caus}. Likewise, perturbations give rise to the mass inflation instability \cite{perts,mass-inf}. Here, perturbations located at the inner horizon induced by external energy fluxes result in an exponential growth of the space-time curvature. The Cauchy horizon is replaced by a (weak) curvature singularity \cite{cauchy-hor}, while the inner horizon collapses, potentially forming a spacelike singularity at a finite time \cite{sing-num}. This does not modify the external structure of the black hole, which is puzzling from the perspective of the no hair theorems. 

Semiclassical effects may play an interesting role when understanding the physics beyond the classical theory \cite{bh-mod,ffnos,blsv,bbcg,abcg,cbcg,abcg2}. However, we are interested in the nonperturbative approach of loop quantum gravity to incorporate quantum corrections (see Refs. \cite{bh-review,bh-book} for a general viewpoint). Concretely, the quantization of uncharged (nonspinning) space-times within the improved dynamics scheme suggested in Refs. \cite{improved,impcova,imp-cont} allowed the derivation of effective geometries with desirable properties. In particular, they agree very well with the exterior geometry of a Schwarszchild black hole for macroscopic masses, but display large corrections close to the would-be singularity, where curvature (including a nonvanishing Ricci tensor) becomes Planckian, and there one finds a transition from a trapped to an anti-trapped region. Charged (nonspinning) space-times have also been studied in Ref. \cite{mato} within loop quantum gravity. Despite there is singularity resolution, we lack of a detailed study of the effective geometries resulting from this quantization. Since they do not adopt an improved dynamics scheme, one should expect they will not show all the desirable properties mentioned above (like Planck-order upper bounds for curvature invariants). This will be the main motivation of this manuscript.

This paper is organized in the following way. In Sec. \ref{sec:class} we describe the Reissner-Nordström space-time in the classical theory. In Sec. \ref{sec:kin} the kinematical aspects of the quantum theory are described and the improved dynamics scheme introduced. Sec. \ref{sec:phys} is devoted to the physical Hilbert space and observables. The main
properties of the effective metric are discussed in Sec. \ref{sec:effec}. Finally, we present the conclusions of the analysis in Sec. \ref{sec:conc}. We also include Appendix \ref{quantization_dynamics} with details about the solutions to the Hamiltonian constraint. 

\section{Reissner-Nordstr\"om space-time: The classical theory}\label{sec:class}

Here we recall the basics of gravity in spherical symmetry \cite{bojo1, bojo2, bojo3,midi,Saeed,Chiou} in  
Ashtekar variables. The gravitational sector consists of two pairs of canonical (gauge-invariant) variables, $E^x, K_x$ and $E^\varphi,K_\varphi$, the triad and the extrinsic curvature, in the radial and transverse directions, respectively. The Poisson brackets between the triad and the extrinsic curvature are
\begin{eqnarray}
\{K_x(x), E^x(x')\}=G\delta (x-x'),\\
\{K_\varphi(x), E^\varphi(x')\}=G\delta (x-x').
\end{eqnarray}
The matter sector is a spherically symmetric electromagnetic field
$A = \Gamma dx+ \Phi dt$ parametrized by two configuration variables $\Gamma$, $\Phi$ and their canonically conjugate momenta,
$P_\Gamma$, $P_\Phi$, respectively. We assume a trivial bundle for the electromagnetic field implying the absence of monopoles. In the canonical treatment it is found that $\Phi$ acts as a Lagrange multiplier, so it can be dropped as a canonical variable (see \cite{mato} for more details). The Poisson brackets  are
\begin{equation}
\{\Gamma(x), P_\Gamma(x')\}=\delta (x-x'),
\end{equation}

The theory has three constraints: the Hamiltonian, diffeomorphism and electromagnetic Gauss law given by 

\begin{eqnarray}
H&=&G^{-1}\left\{-\frac{E^{\varphi}}{2\sqrt{E^{x}}}-2\sqrt{E^{x}}K_{\varphi}K_{x}-\frac{K_{\varphi}^{2}E^{\varphi}}{2\sqrt{E^{x}}}+\frac{\left(\left(E^{x}\right)'\right)^{2}}{8\sqrt{E^{x}}E^{\varphi}} \right. \nonumber\\
&&\left. +\frac{\sqrt{E^{x}}\left(E^{x}\right)''}{2E^{\varphi}}-\frac{\sqrt{E^{x}}\left(E^{x}\right)' \left(E^{\varphi}\right)'}{2\left(E^{\varphi}\right)^{2}}+G\frac{E^{\varphi}P^2_\Gamma}{2(E^{x})^{3/2}} \right\},
\end{eqnarray}

\begin{eqnarray}
C= G^{-1}\left\{ \left(E^{x}\right)'K_{x}-E^{\varphi}\left(K_{\varphi}\right)'-8\pi P_{\phi}\phi' \right\},
\end{eqnarray}

\begin{eqnarray}
\mathcal{G}&=&P'_\Gamma,
\end{eqnarray}
respectively. Here the prime denotes the partial derivative with respect to $x$ and we have set the Immirzi parameter to one. The total Hamiltonian takes the form 
\[
H_{T}=\int dx \left\{ N H + N^x C+\lambda \mathcal{G}\right\},
\]
with $N$ the lapse function, $N^x=g^{xx}N_x$ the shift vector and $\lambda$ the Lagrange multiplier of the Gauss constraint.

In order to make the constraint algebra a true Lie algebra, one can rescale the lapse and the shift functions in the following way

\begin{eqnarray}
\bar{N}^{x}&=&N^{x}+\frac{2NK_{\varphi}\sqrt{E^{x}}}{\left(E^{x}\right)'}\\
\bar{N}&=&\frac{E^{\varphi}N}{\left(E^{x}\right)'}
\end{eqnarray}

so the total Hamiltonian becomes

\[
H_{T}=\int dx \left\{ \bar{N} \left[ \left( \sqrt{E^{x}} \left(1+K_{\varphi}^{2}\right) - \frac{\left(\left(E^{x}\right)'\right)^{2}\sqrt{E^{x}}}{4\left(E^{\varphi}\right)^{2}} -2GM \right)^{'} - \frac{G\left(E^{x}\right)'P_{\Gamma}^{2}}{2 (E^{x})^{3/2}}-\frac{2 G K_{\varphi} \Gamma P'_{\Gamma} }{E^{\varphi}} \right] \right.\nonumber\\
\]
\begin{eqnarray}
&& \left. +\bar{N}^{x}\left[ -(E^{x})'K_{x}+E^{\varphi}(K_{\varphi})'-\Gamma P'_{\Gamma} \right] + \lambda' ( P_{\Gamma} + Q ) \right\}
\end{eqnarray}

The new Hamiltonian constraint (the phase space function multiplying $\bar{N}$ in the above expression) has an Abelian algebra with itself, and the usual algebra with the diffeomorphism constraint and Gauss law (which remain invariant after this redefinition of lapse and shift). The terms $2GM$ and $Q$, with $M$ and $Q$ the Arnowitt-Deser-Misner (ADM) mass and charge, respectively, are
constants of integration that arise from an examination of the theory at spatial infinity.

For static situations we fix the electromagnetic gauge to $\Gamma = 0$, so the Lagrange multiplier $\lambda$ is determined and the Gauss law turns into a strong constraint $P_\Gamma = - Q$. This leads to a total Hamiltonian of the form

\begin{eqnarray}
H_{T}=\int dx \left( \tilde{N}\tilde{H} + \bar{N}^xC_x \right),
\end{eqnarray}
with 
\begin{eqnarray}
C_x=G^{-1}\int dx\left[ -(E^{x})'K_{x}+E^{\varphi}(K_{\varphi})' \right],
\end{eqnarray}

\begin{equation}
    \tilde{H}(\tilde{N})=G^{-1} \int dx \tilde{N} \sqrt{E^{x}} E^{\varphi} \left[ K_{\varphi}^{2} - \frac{\left(\left(E^{x}\right)'\right)^{2}}{4\left(E^{\varphi}\right)^{2}}  + \left(1- \frac{2GM}{\sqrt{E^{x}}} +\frac{G Q^2}{E^{x}} \right) \right],\label{ham const}
\end{equation}
and
\begin{eqnarray}
\tilde{N}=-\frac{1}{E^{\varphi}} \left( \bar{N} \right)'.
\end{eqnarray}

Each of the constraints eliminates one phase space variable per space-time point. In order to have a fully gauge fixed theory, we have to specify the radial coordinate and the spatial slicing (or equivalently the lapse and shift functions). It is important to clarify that for different choices one has diffeomorphically equivalent solutions, so their physical content is the same.
We will restrict to the set of stationary slicings solutions for which $\bar{N}^{x}=0$ and $\tilde{N}=0$ (e.g. $\bar{N}=1/2$). Here, one can easily solve the theory and express the basic phase space variables in terms of two functional parameters $g(x)$, $h(x)$, the ADM mass ($M$) and the charge ($Q$) observables as

\begin{equation}
\begin{array}{cc}
E^{x}(x)=g(x), & (E^{\varphi}(x))^2=\frac{[g'(x)]^2/4}{1+h^2(x)+\frac{G Q^2}{g(x)}-\frac{2GM}{\sqrt{g(x)}}}\\
K_{x}(x)=\frac{[h'(x)]/2}{\sqrt{1+h^2(x)+\frac{G Q^2}{g(x)}-\frac{2GM}{\sqrt{g(x)}}}}, & \;K_{\varphi}=h(x).
\end{array}
\end{equation}

Here $h(x)$ and $g(x)$ (such that $g(x) > 0$ and $g'(x)\neq 0$) are arbitrary functions representing the choice of coordinates for stationary space-times, and requiring that the resulting space-times are asymptotically flat, so $g(x) = x^2 + O(x^{-1})$ and $h(x) = O(x^{-1})$ when $x \rightarrow \infty $. Considering these conditions we have

\[
N^2 = 1+h^2(x)-\frac{G Q^2}{g(x)}-\frac{2GM}{\sqrt{g(x)}},\nonumber \\
\]
\begin{equation}
N^x = \frac{2h(x)\sqrt{g(x)}}{g'(x)} \sqrt{1+h^2(x)+\frac{G Q^2}{g(x)}-\frac{2GM}{\sqrt{g(x)}}}.
\end{equation}

Finally, the space-time metric in spherical symmetry is given by
\begin{equation}\label{eq:class-ds2}
ds^{2}=-(N^{2}-N_xN^x)dt^{2}+2N_xdtdx+\frac{(E^\varphi)^{2}}{\vert E^x\vert}dx^{2}+\vert E^x\vert d\Omega^{2},
\end{equation}
where $d\Omega^2=d\theta^2+\sin^2\theta d\phi^2$ is the metric of the unit round sphere.

\section{KINEMATICS AND IMPROVED DYNAMICS in the QUANTUM THEORY}\label{sec:kin}

We begin the quantization process by recalling the basis of spin network states in one dimension \cite{bojo1, bojo2, bojo3,midi}. Consider graphs $g$ consisting of a collection of edges $e_j$ connecting the vertices $v_j$. One can construct the gravitational sector of the kinematical Hilbert space $\mathcal{H}^{grav}_{kin}$ of the theory, characterized by a basis of states $|\Vec{k}, \Vec{\mu}\rangle$, with $k_j \in \mathbb{Z}$ and $\mu_j \in \mathbb{R}$, the valences of edges $e_j$ and vertices $v_j$, respectively. For convenience, we will restrict our study to spin networks such that $j$ is an integer following a uniform sequence of unit steps in the interval $[-S,-1]\cup[1,S]$, with $S$ finite and arbitrarily large. Therefore each spin network will have a total number of vertices $2S$.\footnote{Note that more general spin networks can be easily considered but the main results of our paper will not change within the semiclassical sector.} 
On this basis, the kinematical operator corresponding to the triad in the radial direction defined on the lattice is
\begin{equation}
\hat{E}^x(x_j) |\Vec{k}, \Vec{\mu}\rangle = \ell^2_{Pl}k_{j}|\Vec{k}, \Vec{\mu}\rangle.
\label{Ex}
\end{equation}

The operator corresponding to the triad in the tangent direction is a well-defined density on vertices 
\begin{equation}
\hat{E}^{\varphi}(x) |\Vec{k}, \Vec{\mu}\rangle = \sum_{v_j}\delta (x-x_j)\ell^2_{Pl}  \mu_{j}|\Vec{k}, \Vec{\mu}\rangle .\label{Ef}
\end{equation}

Point holonomies $\hat{\mathcal{U}}_{\rho_j}:=\widehat{\exp}(i{\rho_j}K_\varphi (x_j))$ of the connection $K_\varphi$ act on verteces $v_j$ in the following way:

\begin{equation}
\hat{\mathcal{U}}_{\rho_j}|\mu_j\rangle=|\mu_j+\rho_j\rangle. \label{holo}
\end{equation}

Note that there are also well-defined operators corresponding to holonomies of the connection component $K_x$, but since we made the Hamiltonian abelian there are no components of the curvature proportional to $K_x$, so we do not need to construct them explicitly.

Now, we will implement the improved dynamics scheme, following similar ideas introduced by Chiou et al. \cite{Chiou} and following \cite{improved}. We start approximating the components of the {\it classical} curvature (of the real connection) by holonomies of finite closed loops along suitable edges generated by the Killing vectors, such that the {\it classical} physical area enclosed by these plaquettes equals the first nonzero eigenvalue of the full LQG area operator $\Delta$ (the so-called area gap). This prescription requires some knowledge from the quantum geometry we have not yet derived, and we hope that at the end of the day we obtain a self-consistent and physically sensible description. For instance, to properly define areas and lengths, we need a metric. Here, we will take the classical metric (in its diagonal gauge) given in Eq. \eqref{eq:class-ds2}, namely, setting $h(x)=0$. Hence, 
\begin{equation}
    g_{\theta\theta}(x)=E^x(x),\quad g_{xx}(x)=\frac{\big([E^x(x)]'\big)^2}{4E^x(x)}\frac{1}{1- \frac{2GM}{\sqrt{E^{x}}} +\frac{G Q^2}{E^{x}}}.
\end{equation}

We now define the areas of closed holonomies that replace the  components of the curvature in the quantum theory on each vertex $v_j$. Let us consider first a plaquette adapted to a 2-sphere, such that its area is equal to $\Delta$. A 2-sphere will have area $4 \pi g_{\theta\theta}(x)=4 \pi E^x(x)$. The plaquette adapted to a 2-sphere must then satisfy 
\begin{equation}\label{eq:imp1}
\Delta= 4 \pi \ell_{Pl}^2 |k_j| \bar{\rho}^2_j,
\end{equation}
where $\ell_{Pl}^2 k_j$ is the eigenvalue of the kinematical operator $\hat{E}^x(x)$ according to (\ref{Ex}). These eigenvalues represent the areas of the spheres of symmetry. Point holonomies (\ref{holo}) of fractional length 
\begin{equation} \label{rhobar}
\bar{\rho}_j = \sqrt{\frac{\Delta}{4 \pi \ell_{Pl}^2 |k_j|}}.
\end{equation}
will produce a shift $|\mu_j\rangle\rightarrow |\mu_j+\bar{\rho}_j\rangle$ in a state which depends on the spectrum of some kinematical operators.
\\
We will adopt a more convenient state labeling $|\nu_j\rangle$ with $\nu_j=\mu_j \sqrt{ \frac{ 4 \pi \ell^2_{Pl} |k_j|}{\Delta}}$. Point holonomies $\hat{\mathcal{U}}_{\bar{\rho}_j}:=\widehat{\exp}(i{\bar{\rho}_j}K_\varphi (x_j))$ have a well-defined and simple action on this new (single-vertex) state basis of $\mathcal{H}^{grav}_{kin}$
\begin{equation}
    \hat{\mathcal{U}}_{\bar{\rho}_j}|\nu_j\rangle=|\nu_j+1\rangle.
\end{equation}

The elements of the basis $|\Vec{k}, \Vec{\nu}\rangle$ are normalized to $\langle\Vec{k}, \Vec{\nu}|\Vec{k}', \Vec{\nu}'\rangle=\delta_{\Vec{k}\Vec{k}'}\delta_{\Vec{\nu}\Vec{\nu}'}$, and kinematical operator (\ref{Ex}) remains unchanged on this basis. The volume operator density is well-adapted on this basis
\begin{equation}
\hat{V}_j|\Vec{k}, \Vec{\nu}\rangle= \sqrt{\frac{\Delta}{4\pi}}\ell^2_{Pl} \nu_{j}|\Vec{k}, \Vec{\nu}\rangle.
\end{equation}

We now consider a basis $|\Vec{k}, \Vec{\nu}, M, Q \rangle$ in $\mathcal{H}_{kin}$ such that
\begin{equation}
\langle\Vec{k}, \Vec{\nu}|\Vec{k}', \Vec{\nu}'\rangle = \delta_{\Vec{k}\Vec{k}'}\delta_{\Vec{\nu}\Vec{\nu}'}\delta(M-M')\delta(Q-Q').
\end{equation}
In addition, and for the very first time in the literature, we will introduce a second improved dynamics condition given by holonomies that form closed plaquettes (anulus) in the the $\theta-x$ and $\varphi-x$ planes. For simplicity, we set the plaquettes in the $\theta = \pi/2$ plane. With respect to the physical metric, infinitesimal  lengths along $x$-direction are given by the norm of the 1-form $(dx)^\mu$, namely, $\sqrt{|g_{\mu\nu}(dx)^\mu(dx)^\nu|}=\sqrt{|g_{xx}|} dx$, where the absolute value has been introduced to make this expression valid even when the 1-form $(dx)^\mu$ becomes timelike (like in the interior of the black hole). Those along the equator are given be the norm of the 1-form $(d\varphi)^\mu$ and have infinitesimal length $\sqrt{g_{\mu\nu}(d\varphi)^\mu(d\varphi)^\nu}=\sqrt{g_{\theta \theta}}d\varphi$. Both in the classical and the quantum theories, we need to specify a choice of a radial coordinate. Following Ref. \cite{gop-imp3}, we define the radial coordinate $\ell_{\rm Pl}^2 k_{j}={\rm sign}(k_j)(x^2_{j}+x_0^2)$, with 
\begin{align} \label{eq:xj}\nonumber
x_j&= (j+1) \delta x \quad {\rm if} \quad  j\in [-S,-1],\\
x_j&= (j-1) \delta x \quad {\rm if} \quad j\in [1,S],
\end{align}
such that $(\delta x/\ell_{Pl})\in \mathbb{N}$ and with $j\neq 0$. Note that  $\ell_{\rm Pl}^2 k_{j=\pm 1}=\pm x_0^2$ fixes the value of $x_0$ to be related with the smallest area of the spheres of symmetry. Its value will be dynamically determined in next Section. In turn, the spectrum of $\left[\left(E^x_j\right)'\right]^2$ can be approximated by $\left(2 \sqrt{x_j^2+\Delta^2/x_0^2}+\delta x\right)^2$. The reason for this approximation is that we did not find a closed-form expression for it. The previous approximated formula agrees very well with the exact spectrum of the above operator, up to corrections of the order $\Delta^2/x_0^2$, which are negligible for macroscopic black holes. 

We now demand that closed plaquettes in the equator of the innermost vertex (where we expect that quantum corrections will be largest) enclose a physical area
\begin{equation}\label{eq:imp2}
2\pi \sqrt{|g_{xx}(x_{j})|} \delta x_{j}  \sqrt{g_{\theta \theta}(x_{j})}\rho_{j}\Big|_{j=1}=
\Delta,
\end{equation}
recalling that $\delta x_j$ is the step of the radial coordinate $x_j$ of the lattice, which here satisfies $\delta x_j=\delta x$ since we choose a uniform lattice spacing,\footnote{This improved dynamics condition involves vertices $j=1$ and $j=2$. One can also obtain a similar result if one instead chooses the vertices $j=-1$ and $j=-2$. Another possibility is to evaluate the improved dynamics condition by choosing the vertices $j=-1$ and $j=1$. Here the spacing of the lattice would be $2\delta x$. In any case, we will get qualitative agreement.} and $2\pi \rho_{j}$ must be understood as a fractional (coordinate) length along the equator in the $\varphi$-direction.\footnote{Note however that a local version of the above improved dynamics condition (valid at all $j$'s) implies a restriction for the allowed sequences of $k_j$ of the spin networks. By itself, is an interesting proposal that deserves to be studied separately.}
Condition \eqref{eq:imp2} takes a complicated form in terms of physical operators. We will study it in the next section.

Besides, in our model, there are global degrees of freedom corresponding to the mass $M$ and the charge $Q$ of the classical space-times. For both of them, we adopt a standard representation as the one already provided in \cite{mato}. Concretely, a complete basis of kinematical states is given by $|\Vec{k},\Vec{\nu}, M, Q \rangle$, normalized such that $\langle \Vec{k}, \Vec{\nu}, M, Q |\Vec{k}', \Vec{\nu}', M', Q' \rangle=\delta_{\Vec{k}\Vec{k}'}\delta_{\Vec{\nu}\Vec{\nu}'}\delta(M-M')\delta(Q-Q')$.
The kinematical operators associated with the mass $\hat{M}$ and the charge $\hat{Q}$ act as multiplicative operators, namely

\begin{equation}
    \hat{M}|\Vec{k}, \Vec{\nu},M, Q \rangle=M|\Vec{k}, \Vec{\nu},M, Q \rangle
\end{equation}
\begin{equation}    
    \hat{Q}|\Vec{k}, \Vec{\nu},M, Q \rangle=Q|\Vec{k}, \Vec{\nu},M, Q \rangle
\end{equation}

Finally, on this kinematical framework, the action of the diffeomorphism and scalar constraints is well defined. Regarding the diffeomorphism constraint, its action is encapsulated in the finite diffeomorphisms (unitary transformations) that map a graph into itself, but moving the positions of the vertices of the one dimensional manifold the have support on. 

On the other hand, following \cite{mato}, the scalar constraint can be defined by means of the quantum operator which is given by
\begin{align}
\nonumber
\hat{H}(\bar{N}) &= \int dx\bar{N} \left( 2\left[\sqrt{\sqrt{|\hat{E}^x|} \left( 1 + \widehat{\frac{\sin^2 (\overline{\rho}_j K_\varphi(x_j))  }{\overline{\rho}_j^2}} + G\frac{\hat Q^2}{ |\hat{E}^x|} \right) - 2G\hat M  }\right]\hat{E}^{\varphi} \right.
\\
\label{total_hamiltonian}
&- \left. \left( \hat{E}^x \right)' (|\hat{E}^x|)^{1/4}\right) .
\end{align}

The operator $\hat{H}(\tilde{N})$ acts on each vertex $\nu_j$ of the kinematical states $|\Vec{k}, \Vec{\nu}, M, Q \rangle$ on $\mathcal{H}_{kin}$ as a first order differential operator in the representation in which holonomies act as multiplicative operators.  On this kinematical Hilbert space, the action of the constraints is well defined and their quantum algebra is free of anomalies. The physical states that are annihilated by this operator will be discussed in the following section.

\section{PHYSICAL SECTOR OF THE THEORY WITHIN THE IMPROVED DYNAMICS}\label{sec:phys}

The previous kinematical description is well adapted to the quantization procedure of \cite{mato}, now within the improved dynamics scheme of \cite{improved}. Here, the scalar constraint is again a differential operator. Its solutions can be obtained as in Ref. \cite{mato} (see Appendix \ref{quantization_dynamics} for details). They represent physical states, where the labels $\nu_j$ play the role of gauge parameters which determine the slicing of the space-time,\footnote{In the classical theory this is equivalent to fix $E^{\varphi}$ with a gauge fixing condition.} and hence are not physical. The space of solutions can be endowed with a physical inner product, and promoted to a Hilbert space. Here, a basis of physical states is $|\Vec{k}, M, Q \rangle$, normalized such that $\langle \Vec{k}, M, Q |\Vec{k}', M', Q' \rangle=\delta_{\Vec{k}\Vec{k}'}\delta(M-M')\delta(Q-Q')$.
The basic physical observables are the mass $\hat{M}$ and the charge $\hat{Q}$, acting on the the physical states in the following way
\begin{equation} 
    \hat{M}|\Vec{k}, M, Q \rangle=M|\Vec{k}, M, Q \rangle
\end{equation} 
\begin{equation}    
    \hat{Q}|\Vec{k}, M, Q \rangle=Q|\Vec{k}, M, Q \rangle
\end{equation}
and a set of observables $\hat{O}$ associated with $\Vec{k}$ and parametrized by a continuous parameter $z \in [-1,1]$

\begin{equation}
    \hat{O}(z)|\Vec{k}, M, Q \rangle = \ell^2_{Pl} k_{Int(Sz)} |\Vec{k}, M, Q \rangle,
\end{equation}
with $2S$ the total number of vertices and Int$(Sz)$ the integer part of $Sz$. The physical observable $\hat{O}$ codifies the (quantized) areas of the spheres of symmetry. As we did with the kinematical states, we will consider spin networks with a finite but large number of vertices. 

In the family of states $|M,Q,\Vec{k}\rangle$, the triad  can be easily represented as physical parametrized observables as
\begin{equation}
\hat{E}^x(x_j) |M,Q,\Vec{k}\rangle = \hat{O}(z(x_j))|M,Q,\Vec{k}\rangle = \ell^2_{Pl}k_{j} |M,Q,\Vec{k}\rangle = {\rm sign}(k_j)(x^2_{j}+x_0^2) |M,Q,\Vec{k}\rangle,
\end{equation}
and its spatial derivative
\begin{equation}
[\hat E^x(x_j)]'|M,Q,\Vec{k}\rangle=\ell_{\rm Pl}^2\frac{k_{j+1}-k_j}{\delta x_j}|M,Q,\Vec{k}\rangle,
\end{equation}
has positive definite spectrum (with its minimum eigenvalue equal to $\delta x$).

From the expression of the Hamiltonian constraint (\ref{ham const}) we can write the square of the triad $\hat{E}^{\varphi}$ as a parametrized observables \cite{bh-book}
\begin{equation}
(\hat{E}^\varphi (x_j))^2=\frac{\left[ (\hat{E}^x(x_j))'\right]^2/4}{1+\frac{\widehat{\sin^2 ({\rho}_j K_\varphi(x_j))} }{{\rho}^2_j} - \frac{2 G \hat M}{\sqrt{|\hat{E}^x(x_j)|}} + \frac{G \hat Q^2}{|\hat{E}^x(x_j)|}},
\end{equation}
with $K_\varphi(x_j)$ playing the role of a collection of parameters, one for each vertex, 
which can depend on $\hat{M}$, $\hat{Q}$ or $\hat{O}(z)$. 

$\hat{E}^\varphi$ must be a well defined self-adjoint operator. Concretely, when $\sin^2 (\overline{\rho}_j K_\varphi(x_j))=1$, it must satisfy $(\hat{E}^\varphi)^2 > 0$, so
\begin{equation}
    1+\frac{1}{\overline{\rho}_j^2} - \frac{2 G M}{\sqrt{|E^x(x_j)|}} + \frac{G Q^2}{|E^x(x_j)|} > 0, \hspace{0.5cm}\forall  \hspace{0.2cm}x_j, M, Q. \label{ineq}
\end{equation}

Condition (\ref{ineq}) leads to a minimum global eigenvalue of $|\hat E^x(x_j)|$, given by $\ell^2_{Pl}|k_{\pm 1}|=x_0^2$, which yields the smallest value of the area of the 2-spheres. At this point of the lattice we expect large quantum effects. If in addition we replace  $\bar{\rho}_0$ by Eq. (\ref{rhobar}), we can obtain the values of $x_0$ fulfilling inequality \eqref{ineq} by analyzing the zeros of
\begin{equation}\label{eq:roots}
\sigma = 1+\frac{4\pi x_0^2}{\Delta}- \frac{r_S}{x_0} + \frac{r_Q^2}{x_0^2} , \end{equation}
for a given mass, $M$, and a given charge, $Q$. Here, we have introduced $r_S=2 G M$ and $r_Q=\sqrt{G}Q$, in order to simplify the notation. 

 If $r_Q$ is small enough compared to $r_S$, we can find two real positive solutions $x^{\pm}_0$ of Eq. \eqref{eq:roots} by setting $\sigma=0$. But, before we focus on their physical consequences, let us note that if we now increase $r_Q^2$, keeping $r_S\gg \ell_{Pl}$, one can see that the two real solutions to Eq. \eqref{eq:roots} converge to only one. Its value is given by
\begin{equation}
    r_Q^{eq} = \frac{3}{8}\left(\frac{\Delta r_S^4}{2\pi}\right)^{1/6}. 
\end{equation}
or, equivalently, 
\begin{equation}
x_0^{eq} = \frac{1}{2}\left(\frac{r_S \Delta}{2 \pi}\right)^{1/3}.
\end{equation}
For larger values of $r_Q$, we have $\sigma >0$, implying that inequality \eqref{ineq} is always satisfied and there is no restriction for the values of $k_j$. 

In this work we will focus on the physical consequences of the limiting case in which $r_Q \ll r_S, r_S \gg \ell_{Pl}$. We leave other regimes to be explored in future works. In the considered limit, the largest solution to \eqref{eq:roots} is given by 
\begin{equation}
\label{xmin}
x_0^{+} = \left(\frac{r_S \Delta}{4 \pi}\right)^{1/3}-\frac{r_Q^2}{3 r_S}+{\cal O}\left[\left(\frac{\ell_P}{r_S}\right)^{4/3}\right]+{\cal O}\left[\frac{r_Q^4}{\Delta^{1/3}r_S^{7/3}}\right]
\end{equation}
while the smaller solution is of order $x_0^{-} \sim r_Q^2 / r_S$.

In principle, all of the eigenvalues of the operator $|\hat{E}^x(x_j)|$ greater than $(x_0^{+})^2$ and smaller than $(x_0^{-})^2$ are allowed. However, in the case in which $x_0^{-}>\ell_{Pl}$, there will be spin networks producing effective geometries that will show an inner core separated from the external region with a strong quantum character. This will require a more delicate analysis that we will leave for a future publication. Hence, we will consider configurations such that $x_0^{-}<\ell_{Pl}$. Note that this actually implies $r_Q^2/r_S<\ell_{Pl}$. Given that we are assuming $r_Q \ll r_S, r_S \gg \ell_P$, the above condition will be likely satisfied. Therefore, the spin networks we will consider here will only have support on eigenvalues of the operator $\hat{E}^x(x_j)$ greater than $(x_0^{+})^2$.

Moreover, the second improved dynamics condition \eqref{eq:imp2} reads
\begin{equation}\label{eq:imp2-x0}
\frac{\sqrt{1+\Delta^2/x_0^4}+\delta x/2x_0^2}{\sqrt{\frac{r_S}{x_0} - 1 - \frac{r_Q^2}{x_0^2}}} \delta x \sqrt{\Delta\pi}=\Delta,
\end{equation}
and in the limit $r_Q \ll r_S, r_S \gg \ell_P$ implies 
\begin{equation}
\delta x = 2 \ell_{Pl} {\rm Int}\left[ \frac{x_0^{+}}{\ell_{Pl}}\right],     
\end{equation}
at leading order. All this fixes the values of the parameters $x_0$ and $\delta x$. It is important to note that condition (\ref{ineq}), which implements the improved dynamics condition \eqref{eq:imp1}, refers to the spectral properties of a parametrized observable in a gauge that in the classical theory allows to cover the maximum extension of the space time and hence allows us to obtain the minimum value for the areas of the spheres of symmetry. We should note that this maximum extension is gauge invariant by construction. On the other hand, once this condition is solved, and the minimum value of $k_j$ is determined, condition \eqref{eq:imp2-x0} refers to the second improved dynamics condition that we must solve for $\delta x$. For it, we choose a diagonal gauge that is well adapted to the calculation of lengths in the radial directions. One could think that this condition is gauge dependent. However, we construct this condition via scalars (norms of two 1-forms), i.e. geometrical quantities that are independent of the gauge one chooses. Therefore, condition \eqref{eq:imp2-x0} is gauge invariant. 

Let us now compare this result  to the uncharged case studied in \cite{improved,imp2021}. The equation \eqref{eq:roots}, when $r_Q = 0$, becomes:
\begin{equation}
\label{eq:roots2}
1+\frac{4\pi x_0^2}{\Delta}- \frac{r_S}{x_0} = \sigma
\end{equation}
Setting $\sigma=0$ we can find one real positive solution to this equation, which in the limit $r_S \gg \ell_{Pl}$ is given by:
\begin{equation}
\label{xmin2}
x_0^{S} = \ell_{Pl} {\rm Int}\left[\frac{1}{\ell_{Pl}}\left(\frac{r_S \Delta}{4 \pi}\right)^{1/3}\right],\quad \delta x^S = 2\ell_{Pl} {\rm Int}\left[\frac{x_0^{S}}{\ell_{Pl}}\right]. 
\end{equation}
For values of $x_j<x_0^{S}$ the quantity $\sigma$ in \eqref{eq:roots2} becomes negative so the inequality \eqref{ineq} is not satisfied. Comparing this result with \eqref{xmin}, we can see that condition \eqref{ineq} leads, in both cases, to a minimum eigenvalue for the operator $\hat{E}^x(x_j)$ both of which differ by a term of order ${\cal O}(r_Q^2 / r_S)$. As we will see in the next section, there is an event horizon located at $\ell_{Pl}\sqrt{|k_j|}=r_S$ in the uncharged case and at $\ell_{Pl}\sqrt{|k_j|}=r_S-r_Q^2 / r_S$ in the charged case. The Cauchy horizon present in the classical Reissner-Nordström black hole is located at a value of $\ell_{Pl}\sqrt{|k_j|}$ which is smaller than $x_0^{+}$ and is therefore inaccessible. From this discussion, we conclude that one of the consequences of adding a small charge to a quantum Schwarzschild black hole is merely that of shifting its event horizon and transition surface locations by a small amount. Besides, as we expected, the global space-time structure will remain qualitatively equivalent to the one of the uncharged black hole.

\section{Reissner-Nordström effective metric}\label{sec:effec}

Let us now construct the line element of the space-time with the purpose of analyzing the physical aspects of the improved dynamics and compare them with the results of \cite{improved}. In order to do this, we are going to use the same slicing as \cite{improved,imp2021}, which uses Eddington-Finkelstein horizon penetrating coordinates such that

\begin{equation}
    \frac{\sin^2{( \widehat { \bar{\rho}_j K_\varphi }(x_j))} }{\bar{\rho}^2_j}=\frac{ \left( -\frac{G\hat{Q}^2}{|\hat{E}^x(x_j)|} + \frac{2G\hat{M}}{\sqrt{|\hat{E}^x(x_j)|}} \right)^2 }{ 1 -\frac{G\hat{Q}^2}{|\hat{E}^x(x_j)|} + \frac{2G\hat{M}}{\sqrt{|\hat{E}^x(x_j)|}} }\label{sin2}.
\end{equation}

The operators corresponding to the metric components are then given by:

\[
    \hat{g}_{tt}(x_j)= 1+\frac{G\hat{Q}^2}{|\hat{E}^x(x_j)|} - \frac{2G\hat{M}}{\sqrt{|\hat{E}^x(x_j)|}}, 
    \hspace{0.4cm} \hat{g}_{xx}(x_j)=\left( \frac{\big(\hat{E}^x(x_j)\big)'}{2\sqrt{|\hat{E}^x(x_j)|}} \right)^2, \hspace{0.4cm} \hat{g}_{\theta\theta}(x_j)=|\hat{E}^x(x_j)|,
\]
\begin{equation}
    \hat{g}_{tx}(x_j)= - \frac{\big(\hat{E}^x(x_j)\big)'}{2\sqrt{|\hat{E}^x(x_j)|}} \sqrt{\frac{2G\hat{M}}{\sqrt{|\hat{E}^x(x_j)|}}-\frac{G\hat{Q}^2}{|\hat{E}^x(x_j)|}} , \hspace{0.4cm} \hat{g}_{\phi\phi}(x_j)=|\hat{E}^x (x_j)|\sin^2{(\theta)}.
\end{equation}
\\
We will now consider a family of quantum states which are sharply peaked in both the mass and charge and are compatible with the restriction to a single spin network (one dimensional lattice). The construction of such states can be done in an analogous way to that of \cite{improved}. They will be peaked on $r_Q \ll r_S$ so that the minimum value for $k_j$ in our spin-network is that corresponding to \eqref{xmin}. We adopt here the approximation $\left[\left(E^x_j\right)'\right]^2$ by $\left(2 \sqrt{x_j^2+\Delta^2/x_0^2}+\delta x\right)^2$ discussed in previous section. An effective space-time metric can then be defined as $g_{\mu \nu}=\langle \hat{g}_{\mu \nu} \rangle$, where the expectation value is computed on the states previously mentioned and where we replace the discrete label $j$ by a continuous dimensionfull coordinate $\ell_{Pl} j \to x\in \mathbb{R}$.\footnote{This approximation is well justified. See Ref. \cite{impcova} for a more detailed treatment of the discrete model. Besides, and for simplicity, we assume $x$ takes values in the whole real line, although we should keep in mind that the fundamental theory has a finite number of vertices and therefore we can only cover a large but finite portion of the space-time.} Moreover, although the expectation values depend on superpositions in the mass $\hat{M}$ and the charge $\hat Q$, we will restrict ourselves to the case in which both $\Delta M$ and $\Delta Q$ are negligible (a discussion of this point can be found in Appendix B of \cite{improved} for the uncharged case). In this limit, the effective metric can be written as $g_{\mu \nu}=^{(0)}g_{\mu \nu}+\ldots$, where ``$\ldots$'' means contributions proportional to $\Delta M^2$,  $\Delta Q^2$ and $\Delta M \Delta Q$ that will be ignored. In total, we obtain the following metric:
\begin{eqnarray}\nonumber
&&^{(0)}ds^2:= \hspace{1pt}^{(0)}g_{\mu \nu}dx^{\mu}dx^{\nu}=-f(x)dt^2+\frac{\left(\sqrt{x^2+\Delta^2/(x_0^+)^2} + x_0^{+}\right)^2} {\left(x^2+(x_0^+)^2\right)}dx^2 \\
&& - \sqrt{1-f(x)} \frac{\left(\sqrt{x^2+\Delta^2/(x_0^+)^2} + x_0^{+}\right)} {\sqrt{\left(x^2+(x_0^+)^2\right)}} dtdx+\left(x^2+(x_0^+)^2\right)d\Omega^2.
\label{eff met}
\end{eqnarray} 
where 
\begin{equation}
    f(x)=1-\frac{r_S}{\sqrt{x^2+(x_0^+)^2}}+\frac{r_Q^2}{x^2+(x_0^+)^2},
\end{equation}
$x_0^+$ is given by Eq. \eqref{xmin} and we have replaced $\delta x$ by Eq. \eqref{eq:imp2-x0}.

\subsection{Curvature of the effective space-time}
In order to compare our results with \cite{improved} we will analyze the properties of the curvature of the effective metric (\ref{eff met}) by computing the Ricci scalar $R_{\mu \nu}g^{\mu \nu}$, the Kretschmann scalar $K=R_{\mu \nu \rho \sigma}R^{\mu \nu \rho \sigma}$ and the Ricci tensor squared $R_{\mu \nu}R^{\mu \nu}$.  Their asymptotic expressions at spatial infinity are, respectively:

\begin{align}
R^2 &=\frac{\left(3 {r_S} {x_0^+}+6 {(x_0^+)}^2-2 \Delta /{(x_0^+)}^2\right)^2}{x^8} +\mathcal{O}\left(\frac{1}{x^{9}}\right)
\\
K &= \frac{12 \left( {r_S}^2+2 {r_S} {x_0^+}+2 (x_0^+)^2\right)}{x^6}+ \mathcal{O} \left( \frac{1}{x^7} \right)
\\
R_{\mu \nu}R^{\mu \nu}&= \frac{6 (x_0^+)^2}{x^6}+\mathcal{O}\left(\frac{1}{x^7}\right)
\end{align}
In the asymptotic region, we should note that the terms involving $Q$ are sub-dominating compared to the main deviations from the classical theory. 

\begin{figure}[h!]
    \centering
    \includegraphics[width=0.79\textwidth]{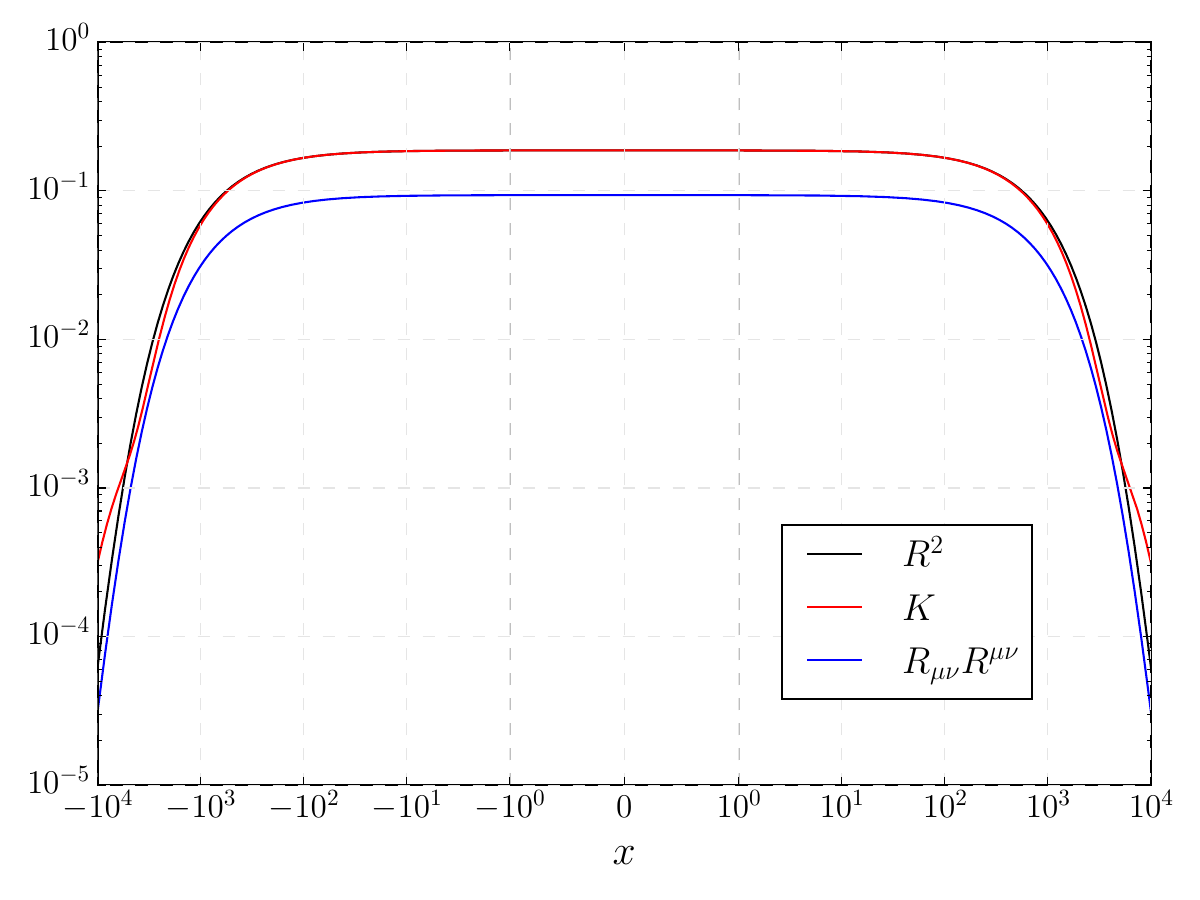}
    \caption{This plot shows the square of the Ricci scalar, the Kretschmann scalar and the Ricci tensor squared for $Q= 1 $ and a choice of $r_S = 2\times10^{10}$ (in natural units).}
    \label{plot1}
\end{figure}

In the most quantum region around $x = x_0$ and in the limit $r_S \gg \ell_{Pl}$ we obtain:
\begin{align}
    R^2 &=\frac{144 \pi^2}{\Delta^2} + \mathcal{O} \left[ r_S^{-2/3} \right]
    \\
    R_{\mu\nu} R^{\mu\nu} &= \frac{72 \pi^2}{ \Delta^2} + \mathcal{O} \left[ r_S^{-2/3}\right]
    \\
    K &= \frac{144 \pi^2}{ \Delta^2} + \mathcal{O} \left[ r_S^{-2/3}\right]
\end{align}
For macroscopic black holes, the curvature invariants reach upper bounds in the most quantum region which are fully determined by the area gap, set in this case as $\Delta=4\sqrt{3} \pi$, where we have set the Immirzi parameter of loop quantum gravity to be $\gamma=1$. In figure \ref{plot1} we have plotted the three curvature invariants, in the most quantum region for small charges ($Q= 1 $ in natural units). The left panel shows the curvature invariants for a choice of the mass parameter corresponding to $r_S = 2\times10^{10}$. 

\subsection{Effective stress-energy tensor}

The effective stress-energy tensor is defined as
\begin{equation}
T_{\mu\nu}:=\frac{1}{8\pi G}G_{\mu\nu}, \label{Tmunu}
\end{equation}
with $G_{\mu\nu}$ the Einstein tensor. It codifies the properties of the effective quantum geometries ($^{(0)}g_{\mu \nu}$) and can be characterized by an effective energy density
$\rho$ and radial and tangential pressures densities, $p_x$ and $p_{\theta}$, respectively. The components of the stress-energy tensor in the exterior region (the region where the effective quantum geometry has a time-like Killing vector $X^{\mu}$) are

\begin{equation}
\rho:=T_{\mu\nu}\frac{X^{\mu}X^{\nu}}{(-X^{\rho}X_{\rho})} ,
\end{equation}

\begin{equation}
p_x:=T_{\mu\nu}\frac{r^{\mu}r^{\nu}}{(r^{\rho}r_{\rho})},
\end{equation}

\begin{equation}
p_{\theta}:=T_{\mu\nu}\frac{\theta^{\mu}\theta^{\nu}}{(\theta^{\rho}\theta{\rho})},
\end{equation}\\
with $r^{\mu}$ the vector field pointing in the radial direction and $\theta^{\mu}$ the vector field pointing in the $\theta$-angular one.
It is worth mentioning that, on the interior region, $X^{\mu}$ becomes space-like while $r^{\mu}$ becomes time-like. This is equivalent to reversing the role of $X^{\mu}$ and $r^{\mu}$ in the previous expressions.
The asymptotic behavior of these quantities at $x\rightarrow \infty$ is given by
{\allowdisplaybreaks
\begin{align}
\rho &=\frac{ r_Q^2+ x_0(2 r_S +3 x_0)}{x^4}+\mathcal{O}\left( \frac{1}{x^5} \right),
\\
p_x &= -\frac{2 x_0}{x^3} + \frac{-r_Q^2 + 3 x_0^2}{x^4} + \mathcal{O} \left( \frac{1}{x^5} \right),
\\
p_{\theta} &= \frac{x_0}{x^3} + \frac{2 r_Q^2 - r_S x_0 - 6 x_0^2}{2 x^4} + \mathcal{O} \left( \frac{1}{x^5} \right) .
\end{align}
}
We can see that the effective stress-energy tensor in the spherical electro-vacuum case falls off sufficiently fast, so the effective metric will come closer to the Minkowski metric at spatial infinity. The energy density and tangential pressure include contributions from $Q^2$. 

Now, in the most quantum region and for macroscopic black holes ($x = 0$, $r_S \gg \ell_{Pl}$), we have:
\begin{align}\nonumber
    \rho &=  \mathcal{O} \left[ r_S^{-2/3} \right],
    \\\nonumber
    p_x &= -\frac{1}{\Delta} + \mathcal{O} \left[ r_S^{-2/3} \right],
    \\
    p_\theta &= -\frac{1}{4\Delta} + \mathcal{O} \left[ r_S^{-2/3} \right].\label{eq:tmunu-bounce}
\end{align}
The components of the stress-energy tensor of a macroscopic black hole reach mass-independent upper bounds completely specified by the area gap $\Delta$.
\begin{figure}[h!]
    \centering
    \includegraphics[width=0.79\textwidth]{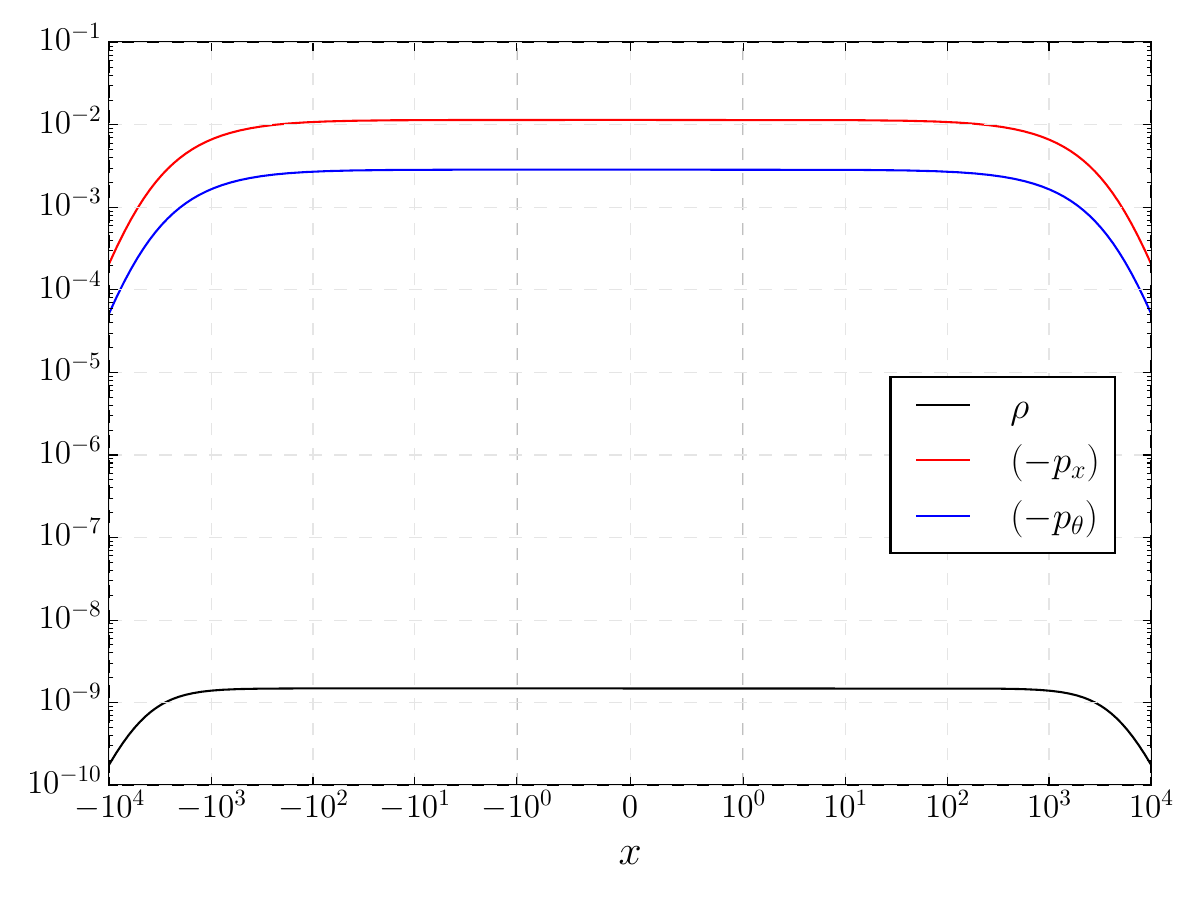}
    \caption{This plot shows the energy and pressure densities for $Q= 1$ and a choice of $r_S = 2\times10^{10}$ (in natural units).}
    \label{plot2}
\end{figure}

As a final comment, let us remember that the singularity theorems (Hawking \& Penrose) are based on the null energy condition. Here, these are given by $\rho + p_x \geq 0$ and $\rho + p_\theta \geq 0$. In the most quantum region, these two conditions are clearly violated from Eq. \eqref{eq:tmunu-bounce}. This is in accordance with the elimination of the singularity by loop quantum gravity. 

\section{Discussion and conclusions}\label{sec:conc}

We have expanded upon the research conducted in \cite{improved} by considering the case of a spherically symmetric black hole with an electric charge. We have restricted our study to the case in which the charge is small compared to the mass of the black hole. After identifying suitable operators for the components of the space-time metric and a suitable family of semiclassical states we derived an effective geometry and studied its properties. 
Most of the results obtained in \cite{improved} seamlessly extend to the charged scenario: i) The effective metric approaches the Reissner-Nordström geometry sufficiently fast at low curvatures; ii) asymptotically, the curvature scalars and the energy and pressure fall off sufficiently fast in an analogous way to that of the uncharged case, except for the energy density which is dominated by $Q^2$; iii) in the most quantum region, the curvature scalars are bounded from above and are at most Planck scale; iv) the effective energy-momentum tensor violates the null energy condition present in the singularity theorems. 
There is, however, a major difference. In \cite{improved} it was found that there is a minimum allowed value for the eigenvalues of the operator $\hat{E}^x$ and this in turn implied the elimination of the singularity for the effective geometry obtained there. In the charged case however we 
found, in the limits $r_Q \ll r_S, ~ r_S \gg \ell_P$ that there is an interval $(k_0^{-},k_0^{+})$ of values that are not allowed so all states have to satisfy $k_j<k_0^{-}, k_j>k_0^{+}$. When constructing the effective metric, however, we wanted to consider states which approximate a smooth geometry and thus we restrict the study to those with a monotonously growing sequence of values of $k_j$ with uniform jumps (of the corresponding radial coordinate) and such that the lowest of them is $k_0 > k_0^{+}$. The resulting semi-classical metric resembles that of the uncharged case constructed in \cite{improved}: It features an event horizon in agreement with the classical theory and a minimum radius in which the curvature is maximum and Planck order. This is a spacelike hypersurface showing a transition from a trapped region to an anti-trapped region, reaching a future (low curvature) white-hole (Cauchy) horizon. This is unlike the classical case in which adding an infinitesimal charge changes the basic structure of the space-time since the classical Reissner-Nordström metric features a (high curvature) Cauchy horizon no matter how small the charge is. In the improved case, this Cauchy horizon is \lq\lq hidden\rq\rq behind an inaccessible region in which quantum operators are not well defined. In \cite{mato} it was mentioned that the discrete nature of space-time could in principle play a role in stabilising the Cauchy horizon of the classical Reissner-Nordström metric. The results of this work seem to point in that direction, since in the effective metric constructed in this work, radiation incoming from $\mathcal{I}^{+}$ can not reach the classical Cauchy horizon since it is located beyond the inaccessible region. However, it will reach the low curvature Cauchy horizon to the future of the anti-trapped region. This deserves a detailed study in the future. 

\section{ACKNOWLEDGMENTS}

This work was supported by program MIA322-CSIC (Universidad de la República, Uruguay), PEDECIBA, Fondo Clemente Estable FCE\_1\_2019\_1\_155865 (ANII). Financial support is provided by the Spanish Government through the projects PID2020-118159GB-C43, PID2019-105943GB-I00 (with FEDER contribution). JO was supported by the “Operative Program FEDER2014-2020 Junta de Andaluc\'ia-Consejer\'ia de Econom\'ia y Conocimiento” under project E-FQM-262-UGR18 by Universidad de Granada during part of the development of this project.

\appendix 

\section{Quantization: Dynamics}
\label{quantization_dynamics}

Starting from the expression \eqref{total_hamiltonian} for the Hamiltonian operator:
\begin{align}
\nonumber
\hat{H}(\bar{N}) &= \int dx\bar{N} \left( 2\left[\sqrt{\sqrt{|\hat{E}^x|} \left( 1 + \widehat{\frac{\sin^2 (\overline{\rho}_j K_\varphi(x_j))  }{\overline{\rho}_j^2}} + G\frac{\hat Q^2}{ |\hat{E}^x|} \right) - 2G\hat M  }\right]\hat{E}^{\varphi} \right.
\\
&- \left. \left( \hat{E}^x \right)' (|\hat{E}^x|)^{1/4}\right) ,
\end{align}
it has a well defined action on kinematical states of the form
\begin{equation}
\label{kinematical_states}
    \left| \Psi \right\rangle = \int \mathrm{d}M \mathrm{d}Q \prod\limits_{v_j} \int_0^{\pi/{\Bar{\rho}}_j} \mathrm{d}K_\varphi (v_j)  \sum_{\Vec{k}} \psi(M,Q,\Vec{k},\Vec{K}_\varphi) \left| M,Q,\Vec{k},\Vec{K}_\varphi \right\rangle.
\end{equation} 
Here, we adopt the ${K}_\varphi$-representation while for ${K}_x$ the usual loop representation.  The main being the simplification of the analysis. Physical states will be constructed out of solutions to the equation $\langle \Psi| \hat{H}^\dagger (\bar{N}) = 0$, where $\langle \Psi|$ are states defined on a dense set of the dual to the kinematical Hilbert space. In the representation we are adopting, we will be dealing with a collection of differential equations rather than finite difference ones in the variables $\nu_j$, which are not solvable in closed form. Since the Hamiltonian operator has the form of a sum of operators acting on different vertices, we may assume
\begin{align}
 \psi(M,Q,\Vec{k},\Vec{K}_\varphi) = \prod\limits_{j} \psi_j (M,Q,k_j,k_{j-1},K_\varphi (v_j)).
\end{align}
It is easy to verify that the action of $\hat{E}^\varphi$ in the loop representation given by \eqref{Ef}, in the connection representation and under the integral becomes simply $\hat{E}^\varphi = -i \ell_P^2 \partial / \partial K_\varphi$. Recalling the action of the operators given by Eqs. \eqref{Ex} and \eqref{holo}, the action of the  Hamiltonian constraint on  states \eqref{kinematical_states} yields 
\begin{align}
\label{equation_physical_states}
    4 i \ell_P^2 \frac{\sqrt{1+m_j^2 \sin^2 (y_j)}}{m_j} \partial_{y_j} \psi_j + \ell_P^2 (k_j-k_{j-1}) \psi_j = 0,
\end{align}
where we have defined
\begin{align}
\nonumber
    y_j &= \bar{\rho}_j K_\varphi(v_j),
    \\
\nonumber
m_j^2 &= \bar{\rho}_j \left( 1 - \frac{2 G M}{\sqrt{\ell_P^2 k_j}} + \frac{G Q^2}{\ell_P^2 k_j} \right),
\end{align}
and with $\bar\rho_j$ given by \eqref{rhobar}.  
Equation \eqref{equation_physical_states} can be solved for $\psi_j$: 
\begin{align}
    \psi_j (M,Q,k_j,k_{j-1},K_\varphi (v_j)) = \exp{\left(\frac{i}{4} m_j (k_j - k_{j-1}) F(\Bar{\rho}_j K_\varphi (v_j) , i m_j )\right)}
\end{align}
with $F$ a two variable function given by
\begin{align}
    F(A,B)=\int_0^A \frac{\mathrm{d}t}{\sqrt{1+B^2 \sin^2 (t)}}
\end{align}
While the steps followed to solve \eqref{equation_physical_states} and the results closely mirror those presented in \cite{mato}, a notable distinction lies in the dependency of the factor $\Bar{\rho}_j$ on $k_j$, whereas in \cite{mato}, it remained a constant.

Physical states are then given by 
\begin{equation}
\label{kinematical_states}
    | \chi \rangle_{phys} = \int \mathrm{d}M \mathrm{d}Q \bigotimes_j\left(\sum_{k_j} \chi(M,Q,k_j) \psi_j (M,Q,k_j,K_{\varphi,j})\left|k_j\right\rangle\right)\otimes\left| M,Q\right\rangle,
\end{equation} 
where the diffeomorphism constrain has been imposed by requiring group averaging. Besides, $\vec{K}_\varphi$ play the role of a collection of parameters that will indicate the choice of slicing. Kinematical states can then be promoted easily to the physical operators (see Sec. \ref{sec:phys}). This is the case of $\hat M$ and $\hat Q$. Other kinematical operators must be promoted as parametrized observables, like $\hat E^x$ and $\hat E^\varphi$.


\begin{thebibliography}{9}

\bibitem{gw16} B. P. Abbott et al., 
Phys. Rev. Lett. {\bf 116}, 061102 (2016).

\bibitem{m87} K. Akiyama et al., 
Astrophys. J. Lett. {\bf 875}, L1 (2019).

\bibitem{penrs} R. Penrose, 
Phys. Rev. Lett. {\bf 14}, pp. 57-59 (1965).

\bibitem{opp-sny} J. Oppenheimer and H. Snyder, 
Phys. Rev. {\bf 56}, pp. 455-459 (1939).

\bibitem{haw-ell} S. W. Hawking and G. F. R. Ellis, {\it The Large Scale Structure of Space-Time}. Cambridge Monographs on Mathematical Physics, Cambridge University Press, 2 2011.

\bibitem{kerr-new} E. T. Newman, E. Couch, K. Chinnapared, A. Exton, A. Prakash, and R. Torrence,
Journal of mathematical physics, {\bf 6},
pp. 918–919 (1965).

\bibitem{kerr-caus} B. Carter, 
Physical Review {\bf 174}, p. 1559 (1968).

\bibitem{perts}  R. H. Price, 
Phys. Rev. D  {\bf 5}, p. 2419 (1972).

\bibitem{mass-inf}  E. Poisson and W. Israel, 
Phys. Rev. Lett. {\bf 63}, p. 1663 (1989).

\bibitem{cauchy-hor} A. Ori, 
Phys. Rev. Lett. {\bf 67}, pp. 789–792 (1991).

\bibitem{sing-num}  P. R. Brady and J. D. Smith, 
Phys. Rev. Lett. {\bf 75}, pp. 1256–1259 (1995).

\bibitem{bh-mod} A. Fabbri and J. Navarro-Salas, {\it Modeling black hole evaporation.} World Scientific, 2005.

\bibitem{ffnos} A. Fabbri, S. Farese, J. Navarro-Salas, G. J. Olmo, and H. Sanchis-Alepuz, 
J. Phys. Conf. Ser. {\bf 33}, pp. 457–462 (2006).

\bibitem{blsv}  C. Barcel\'o, S. Liberati, S. Sonego, and M. Visser, 
Phys. Rev. D {\bf 77}, 044032 (2008). 

\bibitem{bbcg} C. Barcel\'oo, V. Boyanov, R. Carballo-Rubio, and L. J. Garay, 
Class. Quant. Grav. {\bf 36}, 165004 (2019).

\bibitem{abcg} J. Arrechea, C. Barcel\'o, R. Carballo-Rubio, and L. J. Garay, 
Phys. Rev. D, {\bf 101}, 064059 (2020).

\bibitem{cbcg} C. Barcel\'o, V. Boyanov, R. Carballo-Rubio, and L. J. Garay, 
Phys. Rev. D {\bf 102}, 045001 (2020).

\bibitem{abcg2} J. Arrechea, C. Barcel\'o, R. Carballo-Rubio, and L. J. Garay, 
Phys. Rev. D {\bf 104}, 084071 (2021).

\bibitem[]{bh-review} J. Olmedo, Universe {\bf 2} 12 (2016).

\bibitem[]{bh-book} R. Gambini, J. Olmedo, J. Pullin, {\it Handbook of Quantum Gravity}.  Springer, Singapore (2023), edited by C. Bambi, L. Modesto, I. Shapiro.

\bibitem{improved}
R. Gambini, J. Olmedo, J. Pullin, 
Class. Quant. Grav. {\bf 37}, 205012 (2020). arxiv:2006.01513.

\bibitem[]{impcova} R. Gambini, J. Olmedo, J. Pullin, 
Phys. Rev. D {\bf 105}, 026017 (2022). arxiv:2201.01616.

\bibitem{imp-cont}
R. Gambini, J. Olmedo, J. Pullin, 
arxiv:2310.19922

\bibitem{mato}
R. Gambini, E. Mato, J. Pullin, Phys. Rev. D 91, 084006 (2015).

\bibitem{bojo1} M. Bojowald, 
Class. Quant. Grav. {\bf 21}, 3733-3753  (2004).

\bibitem{bojo2} M. Bojowald, and R. Swiderski,
Class. Quant. Grav. {\bf 21}, 4881-4900 (2004).

\bibitem{bojo3} M. Bojowald, and R. Swiderski,
Class. Quant. Grav. {\bf 23}, 2129-2154 (2006).

\bibitem{midi} M. Campiglia, R. Gambini and J. Pullin, Class. Quant. Grav. 24, 3649 (2007) [gr-qc/0703135].

\bibitem{Saeed}
R. Gambini, J. Pullin and S. Rastgoo,
Class. Quant. Grav. \textbf{26}, 215011 (2009).


\bibitem[]{Chiou} D. Chiou, W. Ni and A. Tang, arXiv:1212.1265.
 
\bibitem[]{imp2021} R. Gambini, J. Olmedo, J. Pullin, 
Front. Astron. Space Sci. {\bf 8}, 74 (2021). arxiv:2012.14212

\bibitem[]{gop-imp3} R. Gambini, J. Olmedo, J. Pullin, 
arxiv:2310.19922.

\end{thebibliography}
\end{document}